\documentclass{elsart3}

 \usepackage{graphicx}

\usepackage{amssymb}

\begin{document}

\begin{frontmatter}



\title{Electron-phonon and 
electron-electron interactions \\ in organic field effect transistors}


\author{S. Fratini$^1$, A. F. Morpurgo$^2$ and S. Ciuchi$^3$}
\address{$^1$Institut N\'eel - CNRS \& Universit\'e Joseph Fourier, BP166,
  F-38042 Grenoble, France\\
$^2$ Kavli Institute of Nanoscience, Delft University of Technology, 
Lorentzweg 1, 2628 CJ Delft, The Netherlands \\
$^3$INFM \&
Dipartimento di Fisica, 
Universit\`a dell'Aquila,
via Vetoio, I-67010 Coppito-L'Aquila, Italy}


\begin{abstract}
Recent experiments have demonstrated that the performances of organic
FETs strongly depend on the dielectric properties of the gate
insulator. 
In particular, it has been shown that the temperature dependence of
the mobility evolves from a metallic-like to an insulating behavior upon
increasing the dielectric constant of the gate material.
This phenomenon can be explained in terms of  the formation
of small polarons, due to the polar interaction of 
the charge carriers with the phonons at the organic/dielectric interface. 
Building on this model, the 
possible consequences  of the Coulomb repulsion between the
carriers at high concentrations are analyzed.  
\end{abstract}

\end{frontmatter}

\section{Introduction}

In organic field effect transistors (FETs), charges 
move in a conducting layer located at the interface between 
an organic crystal and  a gate insulator. Commonly used gate insulators such as
SiO$_2$ or Al$_2$O$_3$ are polar materials.
When an
electron moves in the vicinity of such materials, it induces a
long-range polarization that modifies its physical
properties, forming a {\it polaron}. This mechanism is well known in
inorganic semiconductor heterostructures such as GaAs/AlAs.\cite{MoriAndo} 
There, however, due to the weak
polarizabilities and to the extremely low band masses, the polar coupling
only causes a small renormalization  of the electronic properties.

In organic FETs, there are two important differences that can
eventually lead to much more dramatic effects. The first is that
organic crystals have very narrow bandwidths, due to the
weak Van der Waals bonding between the molecules.
The second is that one can use  gate insulators where  the static
($\epsilon_s$) and  high frequency ($\epsilon_\infty$) dielectric
constants  have a large difference in magnitude, i.e.  that are much
more polarizable than GaAs.

Indeed, recent experiments performed on rubrene single crystal FETs 
have shown that the charge carrier mobility at room temperature  decreases upon
increasing the 
dielectric constant of the gate material.\cite{Iosad} 
At the same time, the
temperature dependence of the mobility  evolves from a  metallic-like
to an insulating-like behavior,
\cite{NatMat} which gives strong support to the relevance
of polaron formation.


In this work  
we present a theoretical framework that consistently explains
the observed experimental behavior. 
The model which describes the interaction of the charge carriers with
the ionic polarization at the interface, 
as well as the basic results concerning the formation of small
polarons, are presented in Section 2. The temperature dependence of the
polaronic hopping mobility is calculated in Section 3, where we
briefly  discuss the experimental results of Ref. \cite{NatMat}.
The possible relevance of
the long-range Coulomb repulsion between the carriers at sufficiently
high concentrations is discussed theoretically in Section 4.

\section{Polaron formation}

The interaction between the carriers
and the polar interface is described by the
following Fr\"ohlich Hamiltonian \cite{MoriAndo,Sak,Kirova}
%
\begin{eqnarray}
\label{eq:model}
  H&=& \sum_k \epsilon_kc^+_{k}c_k + \hbar \omega_s \sum_q a^+_q a_q+
\nonumber  \\
  & & + \sum_{kq} M_q c^+_{k+q}c_k (a_q+a^+_{-q}),
\end{eqnarray}
where $c^+_k$, $c_k$, $a^+_q$, $a_q$ are respectively the 
creation and annihilation
operators for carriers, whose  dispersion is $\epsilon_k$,  and
for optical phonons of frequency  $\omega_s$.
The electron-phonon matrix element is defined as
\begin{equation}
  \label{eq:Mq}
  M_q^2=\frac{2\pi \hbar \omega_s e^2}{a S} \frac{\beta}{q} e^{-2qz},
\end{equation}
where 
$a$ is the
lattice spacing in the organic crystal, $S$ is the total surface and
$z$ is the mean distance between the carriers and the interface. The
strength of the electron-phonon coupling is controlled by the
dimensionless parameter
\begin{equation}
  \label{eq:beta}
  \beta=\frac{\epsilon_s-\epsilon_\infty}{(\epsilon_s+\kappa)
(\epsilon_\infty+\kappa)},
\end{equation}
which is a combination of the dielectric constant  of the
organic crystal ($\kappa$) and those of the 
gate material.

The above matrix element was derived in the long-wavelength limit.
A   cut-off at short distances (at $q\sim \pi/a$) is needed in principle 
to account for the discrete
nature of the molecular crystal. \cite{Lepine94,Alex99} 
 However, the finite 
distance $z$ already acts as a short-distance cut-off:
$M_q$ is exponentially reduced at the Brillouin zone
boundaries for $z\gtrsim a$, in which case 
we can  use Eq. (\ref{eq:Mq}) for all practical purposes. A
prescription to deal with discrete lattice effects for $z \lesssim a$ 
is proposed  in the Appendix.

\bigskip
Due the interaction term in 
Eq. (\ref{eq:model}),  the electrons get ``dressed'' by the ionic
polarization of the gate material.
If the interaction is sufficiently strong, the carriers become
self-trapped on individual (or few) organic molecules,
forming  {\it small polarons},  
and hopping-like transport can set in.

The polaron energy in the strong coupling regime is given by 
$E_P= \sum_q M_q^2/\omega_s$.\cite{Mahan}
Integrating this expression 
over the whole reciprocal space leads to the simple result
\begin{equation}
  \label{eq:Ep}
  E_P=\frac{e^2}{2z} \beta.
\end{equation}
The polaron energy turns out to be
independent on the phonon frequency  $\omega_s$, and therefore depends on the
particular gate dielectric only through the parameter $\beta$.
It should be noted that the
exponential decay of the matrix element $M_q$ does not imply an
analogous decay of the polaron energy with distance, but rather
yields a $1/z$ behaviour as can be seen from the above result.

The condition for the formation of a small polaron roughly corresponds to
$E_P\gtrsim t$, where $t$ is the nearest-neighbor transfer integral. 
According to the result Eq. (\ref{eq:Ep}), taking $t=50 meV$
as  representative  for organic crystals and assuming a typical value
$z\sim a$ ($a=7.2\AA$ in rubrene), we see that the
formation of small polarons is expected as soon as $\beta \gtrsim
0.03$. The physical parameters characterizing the different
dielectrics used in Refs. \cite{Iosad,NatMat} are reported in Table I.

\begin{table}
  \centering
  \begin{tabular}{|c|c|c|c|c|}
 \hline \hline
   & $\epsilon_s$ & $\epsilon_\infty$ & $\omega_{LO}  (cm^{-1})$ & $\beta$ \\
\hline
  Ta$_2$O$_5$ & 25& 4.4 & 200-1000 & 0.099 \\
  Al$_2$O$_3$ & 9.4& 3&  400-900& 0.086 \\
  SiO$_2$ & 3.9& 2.1&  400-1240& 0.051 \\
  parylene & 2.9 & 2.56 & 500-1800 & 0.010 \\
  Si$_3$N$_4$ & 5 - 34  & 4.2 & - &0.014-0.11\\
\hline \hline
  \end{tabular}
\caption{Physical properties of the gate dielectrics.
$\omega_{LO}$ represents the typical range of phonon frequencies found
in the literature for
the bulk materials, $\beta$ is  the dimensionless coupling constant. The dielectric constant of rubrene is $\kappa=3$.
The values reported for Si$_3$N$_4$ take into account the
possible formation of the oxide SiO$_x$N$_y$ during  device fabrication.}
\end{table}

\section{Activated mobility}

The mobility of small polarons in the hopping regime 
can be evaluated by mapping the problem onto a two-site
cluster.\cite{LangFirsov} 
In this reduced model, the electrons interact with the 
long-range polarization field created by the interface optical phonons 
via an appropriate collective coordinate 
$Q=\sum_q (1-\cos qa) M_q (a_q+a^+_{-q})$.
 In the adiabatic regime, where the
lattice motion is much slower than the electrons,
the electronic degrees of freedom can be integrated out leading to 
a double-well potential landscape for the  coordinate $Q$.
The mobility in the hopping regime can be 
expressed using standard techniques as
\begin{equation}
  \label{eq:mob}
  \mu_P(T)=\frac{ea^2}{\hbar}\frac{\omega_s}{2\pi T} e^{-\Delta/T},
\end{equation}
where the  Kramers rate is  
determined by the frequency  $\omega_s$ of oscillation  
around the minima of the potential-well, 
\footnote{The 
 prefactor in Eq. (\ref{eq:mob}) is affected by several
microscopic parameters such as the lattice geometry, the 
interaction with multiple phonon modes, the polaron size, dynamical
friction, which can sensibly modify this result.} 
and the activation energy $\Delta$ corresponds to the height of the
barrier.
\footnote{In the above derivation we have implicitly assumed the
adiabatic regime, which is valid when 
$t^2/ (\omega_s \sqrt{E_P T})\gg 1$. 
A formula similar to 
Eq. (\ref{eq:mob}) hods true in the opposite anti-adiabatic
limit. In that case the activation barrier is
$\Delta=\gamma E_P$, and the
prefactor must be replaced by
$(\sqrt{\pi}/3)({t^2}/\sqrt{T^{3}\Delta})$.}
It is related to the polaron energy $E_P$ defined in Eq. (\ref{eq:Ep})
by 
\begin{equation}
  \label{eq:act}
  \Delta=\gamma E_P -t.
\end{equation}
Here $\gamma$ is a parameter which depends on the shape of the
interaction  matrix element:\cite{Mahan}
\begin{equation}
  \label{eq:gamma}
  \gamma=\frac{1}{2}\frac{ \sum_q M_q^2 (1-\cos q a)}
{ \sum_q M_q^2}\simeq\frac{1}{2}-\frac{z}{\sqrt{a^2+4 z^2}}.
\end{equation}
The second equality has been obtained using Eq. (\ref{eq:Mq}), 
which is valid for $z\gtrsim a$.
Note that for a local interaction (Holstein model, $M_q=const$), from
Eq. (\ref{eq:gamma}) one recovers
$\gamma=1/2$. 
A calculation of $\gamma$ including the effects of the discrete
lattice is presented in the Appendix.

\begin{figure}
\centering
\includegraphics[scale=0.7]{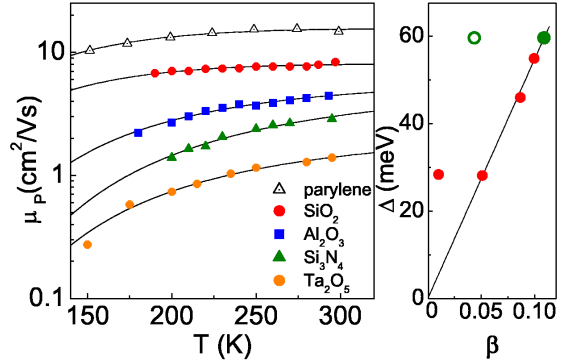}
\caption{(reproduced from Ref. \cite{NatMat}) Left panel: 
polaronic part of the mobility vs. temperature in devices with different gate
dielectrics. Full lines are fits according to
Eq. (\ref{eq:mob}). Right panel: extracted values of the activation
barrier $\Delta$ vs. the dimensionless coupling $\beta$. }
  \label{fig:natmat}
\end{figure}

The temperature dependence of the mobility has been 
measured in Ref.\cite{NatMat} in  devices using the different gate
dielectrics listed in Table I. In practice, the measured $\mu(T)$ includes 
contributions from other
scattering mechanisms, such as the coupling with molecular vibrations
inside the organic crystal. These are entirely responsible for the
mobility $\mu_R(T)$ observed in devices in which the gate insulator is
vacuum, and can be subtracted out from the raw data 
via the Matthiessen rule $\mu^{-1}=\mu_P^{-1}+\mu_R^{-1}$.

In Fig. \ref{fig:natmat} we report the polaronic part of the mobility,
obtained with the above described procedure.
Upon increasing the dielectric constant of the gate insulator
(which amounts to increasing $\beta$), the temperature dependence
evolves towards an insulating-like behavior
that can be fitted
reasonably well with Eq. (\ref{eq:mob}). The values of 
the activation barrier $\Delta$ extracted from the fits are 
reported in the right panel.
Despite the rather limited experimental temperature range,
$\Delta$ is a linear function of $\beta$ for the devices
with the strongest polarizabilities (SiO$_2$, Al$_2$O$_3$,
Ta$_2$O$_5$), as expected from  Eqs. (\ref{eq:Ep}) and (\ref{eq:act}).
Furthermore, 
from the slope of $\Delta$ vs $\beta$, a typical distance $z\sim 3 \AA$
between the carriers and the interface can be estimated, 
which is comparable with the lateral size of a rubrene molecule.
Note that  the effect of $\gamma<1/2$ [cf. Eq. (\ref{eq:gamma})] was not
considered in  Ref.\cite{NatMat}, which led to a higher estimate 
for this distance ($z\sim 6.4 \AA$).

\section{Electron-electron interactions}

Having established that the electron-phonon interactions can be tuned
by changing the polarizability of the gate material, we now
investigate the possibility of controlling the Coulomb interactions
between electrons. It is clear that these are negligible at low carrier
concentrations, i.e.  
in the linear regime  of the $I_{SD}$ vs. $V_G$ curves, where the
conducticity is directly proportional to the carrier density 
(see e.g. Fig. 1 in Ref. \cite{NatMat}). 

However, especially in the devices with strongly
polarizable gate dielectrics, a sizable carrier density  
can in principle be
induced at the highest attainable gate voltages. For example, in 
a device with a
Ta$_2$O$_5$ layer of thickness $d=500nm$, a concentration $x\sim 0.15$
holes per  rubrene molecule can be reached at $V_G=100V$.
At such density, the effects of the electrostatic repulsion
between the carriers cannot be neglected {\it a priori}.
Here we propose a simple theoretical framework to describe
their possible consequences on the device performances.

Let us  start by observing that  
the Coulomb interaction between two electrons in proximity to
a polarizable dielectric is screened by the image charges as
\begin{equation}
  \label{eq:image}
  \varphi(r)=\frac{e^2}{\kappa} \left[ \frac{1}{r} -
\frac{1}{\sqrt{r^2+4z^2}}\frac{
  \epsilon_s-\kappa}{\epsilon_s+\kappa}\right].
\end{equation}
In the following we shall assume that
the average interelectron distance is larger than
the distance $z$ between the electrons and the interface, so that 
Eq. (\ref{eq:image}) 
can be replaced by its asymptotic expression
$2/(\kappa+\epsilon_s)/r$.

In a system of $N$ interacting electrons,
each charge feels instantaneously 
the sum of the repulsive potentials of the
others. Dynamical screening is unlikely 
in the regime under study, where the motion of the charge
carriers is hindered by the formation of small polarons.	
For the same reason, if we focus on a given particle 
while it hops
to its neighboring site, 
the relaxation of the remaining electrons can be neglected to a
first approximation: the collective rearrangement of $N-1$ 
charges necessarily implies
several hopping processes, and therefore occurs on a much longer time scale
than the individual hopping event under consideration.

From the above arguments, if 
the interacting fluid is initially 
at (or sufficiently close to) equilibrium,  it can be expected that 
moving a given particle from a site to its nearest
neighbor will have  
a net energy cost $W$, 
that adds to the polaronic hopping barrier $\Delta$.
In fact, such an energy cost can be
easily included in the two-site model presented in the preceding
Section. It leads to an increased value of the
activation barrier 
$\Delta\to \Delta+W/2$, 
which eventually reduces the carrier mobility 
compared to the result Eq. (\ref{eq:mob}) for independent polarons.
\footnote{This expression is valid  for $W \ll \Delta$. 
More generally, the activation barrier in a biased double-well
  potential is given by $(\Delta+W/4)^2/\Delta$.} 
The mobility  in the presence of
electron-electron interactions 
can thus be expressed as:
\begin{equation}
  \label{eq:mobV}
  \mu(T)=\mu_P(T) e^{-W/2T}.
\end{equation}

To get an order of magnitude 
estimate of this effect, we replace the
actual (uniform) distribution of polarons by a triangular array of point
charges at the same density. 
This approximation is expected to be qualitatively correct
for the following reasons:  first of all,  it
is known\cite{Itoh} that the local correlations in an
interacting charged liquid are very similar to the ones of a
crystallized state. On the other hand, due to the long-range nature of
the interaction  potential Eq. (\ref{eq:image}), 
the result for $W$ should depend only weakly
on the details of the charge distribution. Finally, the proposed 
lattice approximation enforces in a simple way the absence of
relaxation of the electron fluid mentioned above. 

The potential energy of a given
particle in a lattice can be expanded for small
displacements $u$ around its equilibrium position as:
\begin{equation}
  \label{eq:potential}
  \Delta E_{lattice}(u)=\zeta \frac{2}{\kappa+\epsilon_s}
\frac{ e^2}{2R_s^3} u^2.
\end{equation}
Here $R_s$ is the radius of
the Wigner-Seitz disk $n=(\pi R_s)^{-1}$, which is proportional to the
average electron-electron distance, and
$\zeta=0.8$ is a geometrical factor characterizing the triangular
lattice.
In the present approximation, the additional energy cost for hopping
between molecules induced by the electrostatic repulsion between the
carriers is simply given by   $W=\Delta E_{lattice}(a)$. Defining the carrier
concentration according to $a/R_s=\sqrt{\pi x}$ leads to
\begin{equation}
  \label{eq:W}
   W=\zeta \frac{2}{\kappa+\epsilon_s}
 \frac{e^2}{2a} (\pi x)^{3/2} =
 0.31  x^{3/2} eV
\end{equation}
in the case of Ta$_2$O$_5$.
For $x=0.15$, 
we obtain a value $W/2\sim 10 meV$ that is consistently smaller than
the polaronic barrier $\Delta \sim 55 meV$. According to Eq. (\ref{eq:mobV})
the Coulomb interactions between the carriers  can reduce the mobility
by  $30\%$ at room temperature, and even
more at lower temperature. Such 
behavior should be clearly visible as a bending down in the $I_{SD}$
vs $V_G$ curves at the highest attainable values of the gate voltage.
\footnote{It can be argued that such high values of $V_G$ can modify
the charge distribution along $z$, therefore modifying the value of
the activation barrier according to Eqs. (\ref{eq:Ep}), (\ref{eq:act}) and
(\ref{eq:gamma}). 
\cite{Kirova} This effect,
however, should not play a dominant role if the polarons are already
``small''.}
\cite{Iwasa,preparation}

Note that the concentration that can be reached at a given voltage
is proportional to the dielectric constant $\epsilon_s$. As a result,
for devices of equal thickness, the maximum value of 
$W$ increases with $\epsilon_s$ as $\sim
\epsilon_s^{3/2}/(\kappa+\epsilon_s)$. The observation of 
Coulomb interaction effects is therefore more likely in FETs with
highly polarizable gate materials.
 

\section{Concluding remarks}

In this paper we have reviewed a model which describes the interaction
of the charge carriers with the interface phonons in single crystal organic
FETs. Our analysis demonstrates that the 
electron-phonon interaction can be tuned by changing
the dielectric polarizability of the gate material.
This model consistently explains the evolution of the carrier
mobility from a metallic-like to an insulating-like behavior  
that has been recently observed
in Refs. \cite{Iosad,NatMat}: increasing the dielectric polarizability
of the gate insulator results in a crossover from the weak to the
strong coupling regime where
the carriers 
form small polarons, 
which gives rise to
a thermally activated mobility.
Building on this model, the possibility of
revealing experimentally the effects of the long range Coulomb repulsion
between the carriers at high concentrations
has been analyzed.

In conclusion,
although the main interest in organic FETs
comes from their potential applications in "plastic electronics", the 
opportunity of tuning several
parameters such as the carrier density, the electron-phonon and possibly
the electron-electron interactions, makes them an ideal playground for
fundamental physics.

 \appendix

\section{On the lattice cut-off of the electron-phonon interaction.}
\label{App}

Any linear electron-phonon interaction  of the form of Eq. (\ref{eq:model})
can be transformed to
real space as
\begin{equation}
  \label{eq:realpot}
   H_I=\sum_{i,j} g_{ij}c^+_j c_j (a^+_i+a_i)
\end{equation}
where $g_{ij}=\sum_q  M_q e^{-iq(R_{i}-R_j)} $,
$R_i$ being the lattice sites. 
A simple prescription to define a {\it discrete} version of the 
Fr\"ohlich model, appropriate on a lattice, is to
introduce   a short-range cut-off  of the order of the lattice
spacing, which defines a new potential  
$g^{cut}_{ij}$. The correct
periodicity of the matrix element on the Brillouin zone can then be
restored by setting
\begin{equation}\label{periodic}
  \tilde{M}_q= \sum_{R_i} e^{iqR_{i}} g^{cut}_{ij}.
\end{equation}
It is clear that the new matrix element obeys 
$ \tilde{M}_{q+G}=\tilde{M}_q$ with $G$ any reciprocal lattice vector,
which is not the case for $M_q$ of Eq. (\ref{eq:Mq}). 
By taking the continuous limit and letting the cutoff $R_0\to 0$ one
 recovers $\tilde{M}_q=M_q$.

\bigskip

{\it Fr\"ohlich interaction in 3D.}
The essence of the Fr\"ohlich model is the interaction of a charge
(monopole) with a field of dipoles in an ionic material (+ and -
ions). Correspondingly, the usual polar electron-phonon interaction
in 3 dimensions, $M_q^{3D}=M_0/q$,  scales in real space as $1/R^2$,
namely 
\begin{equation}
  \label{eq:3D}
  g_{ij}^{3D}=\frac{M_0}{2\pi^2 R_{ij}^2}.
\end{equation}
The simplest way to introduce a lattice cut-off
is to flatten off the potential at
short range by setting $g_{ii}=g_{i,i+1}$. Alternatively, one can
define 
\begin{equation}
  \label{eq:3Dcut}
  g_{ij}^{3D,cut}=\frac{M_0}{2\pi^2 (R_{ij}^2+R_0^2)}
\end{equation} 
with $R_0$ of the order of the lattice spacing $a$. 
Transforming back $ g_{ij}^{3D,cut}$ to  reciprocal space 
leads to the familiar Yukawa form.

A physically sound choice for the cutoff length $R_0$ in Eq. (\ref{eq:3Dcut})
is to set $E_P$ 
equal to the energy of
self-interaction  of a uniform charge distributed on a cube of side
$a$,\cite{Lepine94} which  yields $R_0\simeq 0.4 a$.
However, the polarizability of the medium at short distances is
certainly less than
what is predicted by  continuum theory,
so that the above estimate
should be considered as a lower bound for $R_0$.

Note that the potential Eq. (\ref{eq:3Dcut}) 
looks essentially similar to the one proposed in
Ref. \cite{Alex99} to describe the interaction with the apical oxygens
in the superconducting 
cuprates, except for the different decay at long distances,
reflecting a different physical origin of the interaction (in
that case, $g_{ij}\sim 1/R_{ij}^3$). As for the short distance
cut-off, the authors 
of Ref.  \cite{Alex99} use $R_0=a$. 

\bigskip
{\it Polar interaction at interfaces.}
The interaction term of Eq. (\ref{eq:Mq}), that we rewrite here as
$M_q=M_0e^{-qz}/\sqrt{q}$, already has a short distance cut-off
determined by the distance $z$ from the interface. 
Indeed, transforming to real space yields
 the following Hypergeometric function
\begin{equation}\label{eq:hyper}
  g_{ij}=\frac{M_0 }{4\sqrt{\pi}z^{3/2}} \; {}_2F_1(3/4,5/4,1,-R_{ij}^2/z^2),
\end{equation}
that scales as $1/R_{ij}^{3/2}$
for large $R_{ij}\gg z$, but tends to a
constant $g_{ii}=1/4\sqrt{\pi}z^{3/2}$ for $R_{ij}\ll z$. Again, a  
lattice cut-off can be introduced in this expression 
either by  flattening off the local
term $g_{ii}$ or by replacing  $z\to \max(z,a)$.

As a final remark, to avoid dealing with the special function ${}_2F_1 $, 
a simpler model can be introduced that reproduces both its asymptotic
limits (but slightly underestimates the polaron energy). It reads
\begin{equation}
  \label{eq:simpler}
  g_{ij}=\frac{ c M_0}{(R_{ij}^2+R_0^2)^{3/4}}
\end{equation}
where  $c=1/[4\Gamma(1/4)\Gamma(5/4)]
\simeq 0.076$ and  $R_0=(4\sqrt{\pi}c)^{2/3} z\simeq 0.66
z$.

The polaron energy  in this model
can be evaluated in real space as $E_P=\sum_i
g_{ij}^2/\omega_s$. Performing the discrete sum 
 on a square lattice in the case $z=a$ 
gives a result which is about $30 \%$
smaller than the result of the continuous approximation. 
On the other hand, the value of the parameter $\gamma$
that determines the activation barrier increases by almost a factor
$2$, leading to an overall enhancement of the barrier. Although these
numbers are quite sensitive to the choice of the cut-off prescription,
they confirm that the continuous approximation used in the text is
reasonably accurate for $z\gtrsim a$.


\begin{thebibliography}{00}

\bibitem{MoriAndo} N. Mori, T. Ando, Phys. Rev. B 40, 6175 (1989)

\bibitem{Iosad} A. F. Stassen, R. W. I. de Boer, N. N. Iosad,
  A. F. Morpurgo, Appl. Phys. Lett. 85, 3899 (2004)
\bibitem{NatMat} I. N. Hulea et al., Nature Materials 5, 982 (2006)

\bibitem{Sak} J. Sak, Phys. Rev. B 6, 3981 (1972)
\bibitem{Kirova} N. Kirova, M. N. Bussac, Phys. Rev. B 68, 235312 (2003)
\bibitem{Mahan} G. D. Mahan,  
{\it Many-Particle Physics}, 2nd ed. (Plenum Press,
New York, 1990)
\bibitem{Lepine94} Y. L\'epine,  J. Phys. Cond. Matter 6, 6611 (1994)
\bibitem{Alex99} A. S. Alexandrov and P. E. Kornilovitch,
  Phys. Rev. Lett. 82, 807 (1999)
\bibitem{LangFirsov}I. G. Lang, Yu. A. Firsov, Sov. Phys. Solid State 9, 2701 (1968)

\bibitem{Itoh} N. Itoh, S. Ichimaru, S. Nagano, Phys. Rev. B 17, 2862 (1978)

\bibitem{Iwasa} Y. Iwasa,  talk at the 2007 CERC Workshop, Tokyo May 2007
\bibitem{preparation} S. Fratini, H. Xie, I. N. Hulea, S. Ciuchi,
 and A. F. Morpurgo, in preparation. 






\end{thebibliography}
\end{document}